\documentclass[aps,amssymb,prl,twocolumn]{revtex4}
\usepackage{graphics}

\def\d {{\rm d}}

\begin{document}

\title{Out-of-equilibrium relaxation of a time-dependent effective temperature}
\author{Ana\"el Lema\^{\i}tre}
\affiliation{
Department of physics, University of California, Santa Barbara, California 93106, U.S.A.}
\affiliation{
L.M.D.H. - Universite Paris VI, UMR 7603, 4 place Jussieu - case 86, 75005  Paris - France}
\date{\today}

\begin{abstract}
Constitutive equations are proposed for the relaxation of a glassy material 
in contact with a thermal reservoir.
The dynamics of a single state variable, identified as an effective 
temperature accounts for the emergence of glassy behavior at low bath temperature. 
Memory effect in cooling/reheating
experiments are discussed along with several aspects of the response to shear:
aging of the response to step strains, response to oscillatory forcing, and steady rheology.
\end{abstract}


\maketitle

The purpose of the present work is to introduce a limited set of kinetic equations
which describe the out-of-equilibrium relaxation of a structural glass 
and its response to shear deformation.
It was originally motivated by recent theories
for the plasticity of amorphous solids,~\cite{falk98,falk00}
in an attempt to incorporate glassy relaxation 
at an elementary level.~\cite{lemaitre02b,lemaitre02c}
A quite simple picture emerges, which accounts for important properties 
of glassy materials,
while its premises may hold for general classes of complex fluids;~\cite{larson99}
it echoes early theories of structural 
relaxation~\cite{turnbull61,adam65,narayanaswamy71,moynihan76}
in a much more systematic framework,
and provides theoretical grounding for phenomenological 
rate-and-state equations.~\cite{derec01}

We consider, in this study, a piece of material at pressure $P$, 
in contact with a thermal reservoir at temperature $T$. 
We sort the material into subsystems~\cite{adam65} (or zones) of, say, $z$ molecules,
and characterize structural disorder by the distribution $\rho$ of volumes,
$v_i\in[v_0,\infty]$, or enthalpies ($h_i=P v_i$) of these molecular configurations.
($v_0$ is a lower bound imposed by excluded volume effects.)
At low $T$, the distribution $\rho$ evolves slowly, as opposed to the vibrations of
the molecular structure which quickly equilibrate with the thermal bath.
The evolution of $\rho$ is expected to result from transitions between metastable 
basins in phase space, and these transitions are supposed to be triggered by
conformational changes of molecular subsystems (rearrangements).
The typical size of a rearranging zones fixes $z$, which
is here supposed to be independent of $T$ and $P$.
At all times, molecular subsystems are in mechanical contact
with neighboring subsystems, hence likely to exchange energy among themselves:
very small displacements of the average positions of the molecules are sufficient 
to ensure these volume (or enthalpy) transfers.
Mediated by acoustic modes (phonons), these transfers are thus expected to be fast, 
collective, processes, in contrast with rearrangements which, 
at low $T$, are expected to be rare, space-localized events.

This observation in the basis of the present work: it permits to separate
these two thermodynamic processes
(equilibration with the bath {\it v.s.\/} transfers between subsystems)
and assume that {\it spatial\/} enthalpy fluctuations 
equilibrate on timescales at which the system is {\it not\/}
in contact with the thermal reservoir. At all time, approximation is made that 
spatial fluctuations are in ``adiabatic equilibrium'', 
characterized by a ($P$,$H$)-ensemble: 
the distribution $\rho$ is determined {\it at every single moment\/}
by statistical inference: it maximizes entropy under the single constraint
that $H=\int\d h\,\rho(h)\,h$ is known, whence $\rho(h) = \frac{1}{Z}\,\exp(-\lambda h)$,
with the partition function
$Z = \int_{h_0=P v_0}^{\infty}\,\d h \exp(-\lambda\,h) = (1/\lambda)\,\exp(-\lambda\,h_0)$,
and where $\lambda$ is the Lagrange multiplier that enforces adiabaticity.
$\Theta=1/(k \lambda)$ is an effective temperature, 
but differs from the temperature of the thermal reservoir, 
unless the system has reached thermal equilibrium.
This effective temperature is a generalization of Edwards':~\cite{edwards94}
I made no assumption here that the system is exploring all configurations 
of some ergodic component at any relevant timescale but, rather, that,
by being very large, it samples, at any single time,
all possible configurations allowed by macroscopic constraints.
At all time, $k\,\Theta(t) = H(t)-P v_0$.~\footnote{$k\,\Theta(t)= P(V-v_0)$
is proportional to a Van der Walls' free-volume:~\cite{gallavottibook99}
in light of the following discussion, free-volume activation~\cite{turnbull61}
and effective temperature,~\cite{edwards94}
appear to be closelly related concepts.}

The previous Ansatz for $\rho$ serves here as a basis to investigate the relaxation
of the system towards equilibrium. 
At any time, $H$ (or $\Theta$) suffice to characterize
the internal state of the material.~\footnote{
The limit we consider is in some sense, ``opposite'' to the mean-field approximation, 
in which case every variable $v_i$ is directly coupled with the average $\bar v$, 
whence the full distribution $\rho(v_i-\bar v)$ becomes the dynamical variable. 
We see that the mechanical contacts between subsystems may 
(below some critical dimension)
drastically reduce the dimensionality of the relevant dynamics.}
Their dynamics is specified by energy conservation: it results from the balance 
between thermal exchanges with the reservoir (heat) and the work of external forces.
Estimating the rate of thermal exchanges with the bath is a difficult task; 
it requires, in principle, to determine all probabilities $p(h\to h')$ 
for the reconformations of molecular subsystems, leading to:
\begin{equation}
\label{eqn:chi1}
\frac{\d H}{\d t} =
\int \d h \d h'\,p(h\to h')\,\rho(h)\,(h'-h) + W^{\rm ext}
\quad.
\end{equation}
I argue that a generic form of this equation arises from 
the study of dominant contributions to the integral term.
Then, I show that the resulting equation accounts for the emergence of slow 
relaxation and glassy behavior at low $T$, with a marked cooling rate dependence 
of the glass transition point. A study of the response to shear in various experimental 
set-up completes my presentation.

Rearrangements correspond to elementary contacts with the thermal reservoir: a fluctuation $\delta h$ 
of enthalpy requires a transfer of heat $\delta q=-\delta h$ 
from the thermal reservoir, hence
occurs with probability $\exp(\delta S/k)=\exp(-\delta h/(kT))$.
These thermally activated processes are very sensitive to the external constraints imposed
on a molecular subsystem by its surroundings: 
for low $v$, the reconformation of the molecules may require to deform their ``cage''
and pay a consequent price in elastic energy; for large $v$, molecules may move freely.
Suppose, for the sake of simplicity, that the transitions $h\to h'$
are controlled by some energy barrier $h_b$ which is a simple
function of the volume $v$.
Obviously, $h_b(v)$ must be non-increasing, hence
there exist an {\em activation volume\/} $v_a$, such that $h_b(v_a)=P v_a\equiv h_a$;
beyond this point ($v>v_a$) the notion of barrier breaks down.
Two different families of reactional pathways must be identified:
when $h,h'<h_b$, transitions are activated, and transition state theory indicates that
the rates must be written: $p(h\to h') = \nu\exp(-(h_b-h)/(kT))$;
otherwise, there is no barrier {\it per se\/}
and Monte-Carlo weights read: $p(h\to h') = \nu\min(1,\exp(-(h'-h)/(kT)))$.
These transition probabilities verify detailled balance and are
continuous functions of either $h$ or $h'$ when they cross $h_a$.
$\nu$ is an update frequency, the temperature dependence of which 
might be neglected as it brings only minimal corrections.
The calculation of the rhs of equation~(\ref{eqn:chi1}) is further simplified by 
allowing only transitions such that $h-h'=\pm\delta q$;
the integral is separated into
the sum $I_1$ over $h,h'\in[h_0,h_a]$, and $I_2$ over the complementary domain of integration.
It appears that $I_1$ (resp. $I_2$) is of order $O((\Theta - T)^2)$ (resp. $O(\Theta-T)$) close to equilibrium,
and is proportional to a factor of the form $\exp(-A/(kT))$ (resp. $\exp(-A'/(k\,\Theta))$) when $\Theta>>T$.
In both these limits, $I_1$ is dominated by $I_2$:
the dynamics of $\Theta$ is controlled by the transition pathways
(found with a frequency $\exp(-P (v_a-v_0)/(k\,\Theta))$) which do not involve the crossing of energy barriers.
It is therefore sufficient to restrict our discussion to the situation in which $I_1\to 0$.
To fix ideas, I assume that $h_b(v)\to\infty$ below $v_a$ and
drops sharply at that point. (Such steep decay of $h_b$ is expected, for example, 
in the case of hard-sphere materials.)

After integration one gets from equation~(\ref{eqn:chi1}):
\begin{equation}
k\,\dot\Theta =
E_1\,\exp\left[-\frac{\Delta h}{k\,\Theta}\right]\,
\left(\exp\left(\frac{\delta q}{k \Theta}-\frac{\delta q}{k T}\right)-1\right)
+\sigma\,\dot\gamma
\quad,
\label{eqn:chi}
\end{equation}
with $E_1=\nu\,\delta q$ and $\Delta h=P (v_a-v_0)$.
I anticipate here on the second part of this paper where the deformation of a material in  
a pure shear geometry will be considered: $\sigma$ is the (deviatoric) shear stress, 
$\dot\gamma$ the plastic strain rate, whence $W^{\rm ext}=\sigma\dot\gamma$.
The results presented here do not depend qualitativelly on further dependencies
of $E_1$ on $T$ or $\Theta$, which may, however, bring logarithmic corrections.

In the absence of external stress ($\sigma=0$),
equation~(\ref{eqn:chi}) admits thermodynamic equilibrium, $\Theta=T$, 
as its single fixed point.
The timescale of the relaxation towards equilibrium increases 
with decreasing $T$, and at low $T$ the system may remain out-of-equilibrium
on any observable timescale by the mere fact that $\Theta$ is small (or tries to).
When $\Theta>> T$, equation~(\ref{eqn:chi}) is dominated by:
\begin{equation}
k\,\dot\Theta = -E_1\,\exp\left[-\frac{\Delta h}{k\,\Theta}\right]+\sigma\,\dot\gamma
\quad.
\label{eqn:chi:zero}
\end{equation}
In this limit, the long-time relaxation of the variable $\Theta$
is easily obtained by integrating~(\ref{eqn:chi:zero}) (with $\dot\gamma\sigma=0$): 
$\int_{\Theta_0}^{\Theta(t)} \exp\left[{\Delta h}/{k\,\Theta}\right] k\,\d \Theta = - E_1\,t$,
and keeping only the dominant contribution: 
$k\,\Theta(t)\simeq \frac{\Delta h}{\log\left(E_1\,t/\Delta h\right)}$.
For finite $T$, this process dominates so long as
$t << \tau_T=\Delta h\,\exp\left[\Delta h/(kT)\right]/E_1$:
$\tau_T$ diverges in the limit $T\to0$.

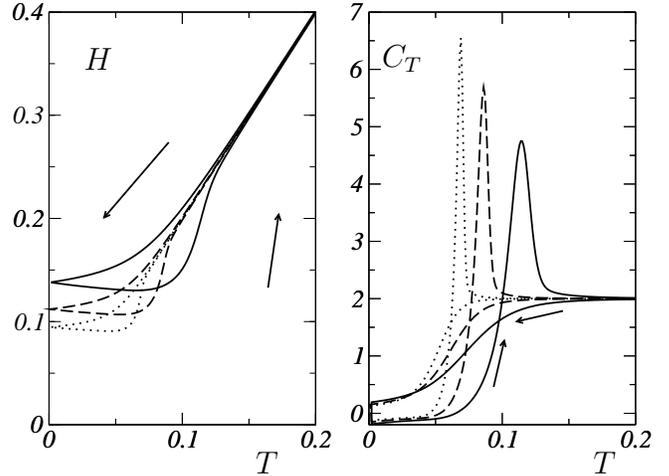
\begin{figure}
\unitlength = 0.005\textwidth
\begin{picture}(100,70)(0,0)
\put(2,2){\resizebox{95\unitlength}{!}{\includegraphics{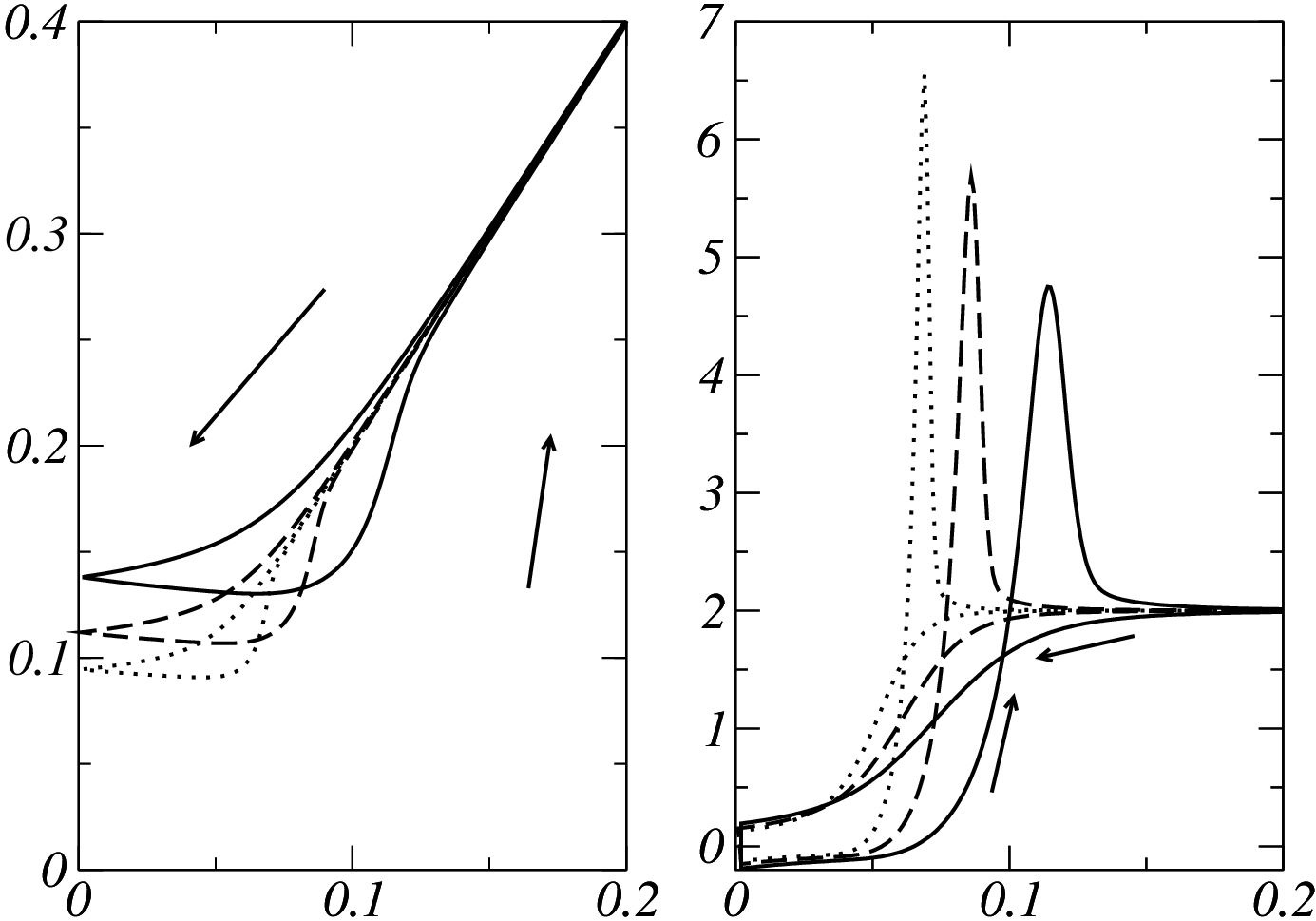}}}
\put(15,60){\makebox(0,0){\large $H$}}
\put(40,0){\makebox(0,0){\large $T$}}
\put(60,60){\makebox(0,0){\large $C_T$}}
\put(90,0){\makebox(0,0){\large $T$}}
\end{picture}
\caption{\label{fig:cooling}
Energy (left) and calorific capacity (right) as a function of temperature during 
cooling and reheating protocol, between $k T_{\rm max}= 0.5$ and $k T_{\rm min}= 0.001$,
for cooling rates, $10^{-3}/k$ (solid line), 
$10^{-4}/k$ (dashed line), $10^{-5}/k$ (dotted line).
}
\end{figure}

An important property of the glass transition is  that 
the observed transition point depends on the cooling rate.~\cite{moynihan76,larson99}
Consider the following experiment:
enthalpy is monitored during a steady cooling
from an initial temperature $T_{\rm max}$; after reaching $T_{\rm min}$,
the temperature is increased at the same rate back to $T_{\rm max}$.
This procedure is reproduced here by
integrating equation~(\ref{eqn:chi}) for different cooling/reheating rates.
The resulting dynamics are portrayed figure~(\ref{fig:cooling}) (left), where
energy is shown as a function of $T$.
Slowing down is seen as energy does not reach equilibrium
at low temperature; upon reheating, energy does not go back along the same curve.
As in the experiments, heat capacity is then extracted from this data according
to the formula, $C_T=\d H/\d T$;
the result is displayed figure~(\ref{fig:cooling}) (right).
The large increase of $C_T$ observed upon reheating marks a fictive 
glass transition point, which clearly depends on the cooling rate.

The rest of this paper is devoted to the study of a material 
under shear. 
Like thermal equilibration, the deformation of a material under an applied
shear stress is expected to proceed by space-localized rearrangements.~\cite{falk98}
However, the transition pathways that contributes respectivelly
to elementary shear and elementary thermal equilibration processes, 
are not necessarily identical at the molecular level.
Therefore they may, in general, display different activation enthalpies.
Without more knowledge, we can expect the response 
to small shear stresses to be of the form:~\footnote{
I do not consider, in this work, dynamics of shear transformation zones.~\cite{falk98}
}
\begin{equation}
\dot\gamma = E_0\,\exp\left[-\frac{\Delta h'}{k\,\Theta}\right]\,\sigma
\label{eqn:gammadot}
\quad.
\end{equation}
The update frequency $E_0$ normalizes the frequency of elementary vibrations,
and incorporates corrections specific to this mode of deformation.
Any dependency of $E_0$ on $\Theta$ or $T$ is neglected.
($E_0/E_1$ and $\Delta h/\Delta h'$ are expected to be of order 1.)

With equations~(\ref{eqn:gammadot}) and~(\ref{eqn:chi}) in hands
I now study three experimental protocols:
aging in the stress-linear response to step strains,
response to oscillatory forcing, and rheology of steady deformation.

Aging is characterized by monitoring the response of a material to a perturbation
at various times after an initial quench from a high temperature 
equilibrium state.~\cite{struik76,larson99}
The procedure is as follows: At time $t=0$, the system is suddenly quenched from
an equilibrium state ($\Theta_0=T_0$) to low temperature.
Relaxational dynamics take place from then on, and we study the situation 
when $\Theta>>T$: equation~(\ref{eqn:chi:zero}) leads at long time to a 
time-logarithmic decay of enthalpy.
At time $t_w$, a small strain $\delta\epsilon$ is applied; it provokes 
a quasi-instantaneous
elastic response $\sigma=2\mu\delta\epsilon$, followed by a slow occurence 
of plastic deformations and stress release as: 
$\dot\sigma=-2\mu\dot\gamma$. For small $\delta\epsilon$, the term $\dot\gamma\sigma$ in 
equation~(\ref{eqn:chi:zero}) is negligible at all times.
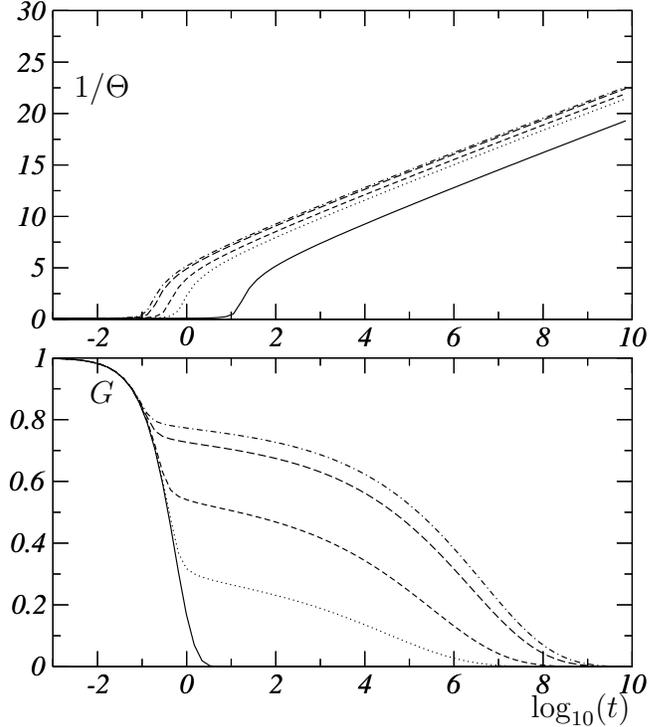
\begin{figure}
\begin{center}
\unitlength = 0.005\textwidth
\begin{picture}(100,100)(0,0)
\put(0,0){\resizebox{95\unitlength}{!}{\includegraphics{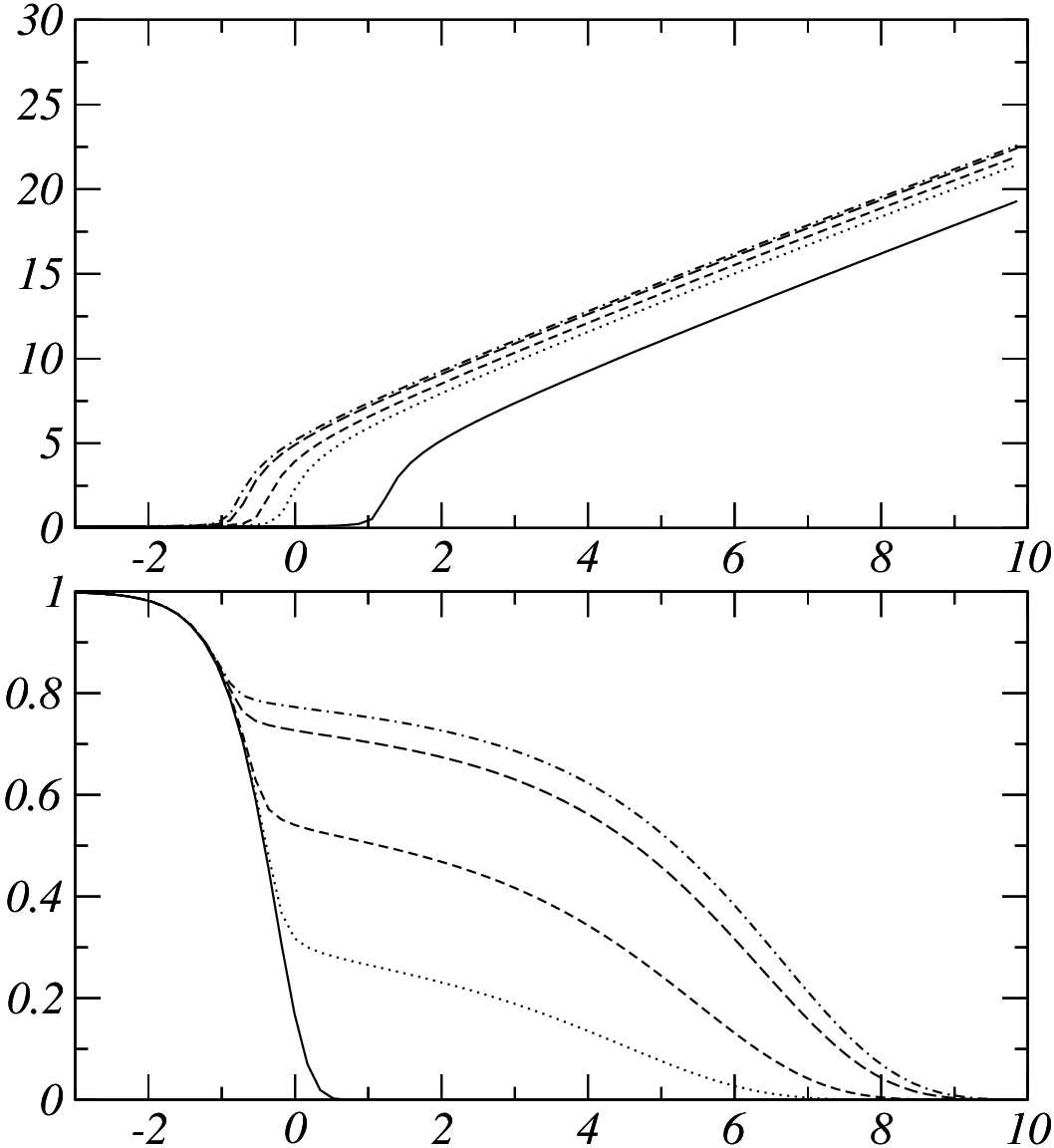}}}
\put(14,90){\makebox(0,0){\large $1/\Theta$}}
\put(14,45){\makebox(0,0){\large $G$}}
\put(85,-2){\makebox(0,0){\large $\log_{10}(t)$}}
\end{picture}
\end{center}
\caption{\label{fig:relax1}
Parameters are $E_0=\mu=1$, $\Delta h=1.5$, $\Delta h'=1$, $\Theta(0)=10$, and $t_w=0$.
Top: $1/\Theta$ as a function of $\log_{10}(t)$
for $E_1 = 1, 20, 40, 80, 100$ from right to left.
Bottom: relaxation spectra with $E_1$ increasing from left to right.
}
\end{figure}

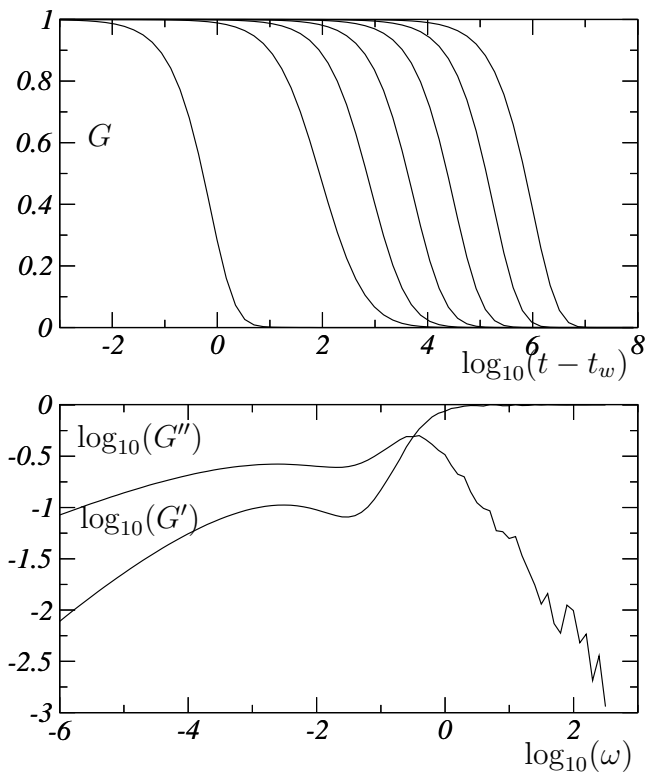
\begin{figure}
\begin{center}
\unitlength = 0.005\textwidth
\begin{picture}(100,110)(0,0)
\put(0,0){\resizebox{95\unitlength}{!}{\includegraphics{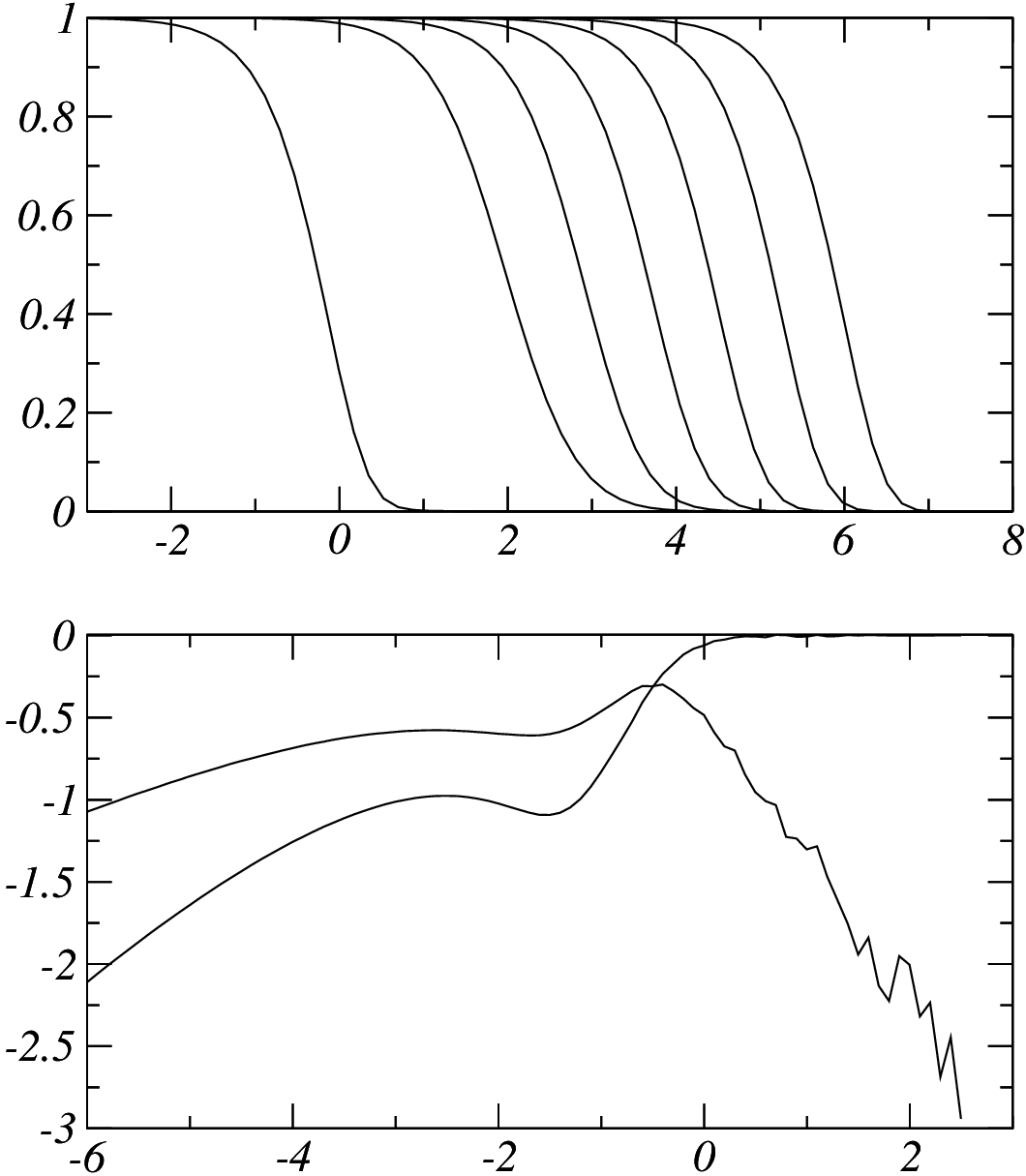}}}
\put(14,90){\makebox(0,0){\large $G$}}
\put(20,45){\makebox(0,0){\large $\log_{10}(G'')$}}
\put(20,33){\makebox(0,0){\large $\log_{10}(G')$}}
\put(80,56){\makebox(0,0){\large $\log_{10}(t-t_w)$}}
\put(85,-2){\makebox(0,0){\large $\log_{10}(\omega)$}}
\end{picture}
\end{center}
\caption{\label{fig:relax}
Top: Relaxation modulus for $E_0=E_1=\mu=1$, $\Delta h=1.5$, $\Delta h'=1$, $\Theta(0)=10$, 
and $t_w=10,10^2,10^3,10^4,10^5,10^6,10^7$.
Bottom: Normalized frequency spectra for $\mu=100$, $E_0=E_1=0.01$, $\Delta h=1.5$, $\Delta h'=1$, $t_w=10$, $\Theta(0)=10$.
}
\end{figure}

{\it At short times\/}, stress decays exponentially, with a timescale,
$\tau_\Theta = (1/(2\mu\,E_0))\,\exp\left[{\Delta h'}/(k\Theta_w)\right]$,
which depends only on $\Theta_w=\Theta(t_w)$.
{\it At long-times\/}, stress release is controlled by the slow relaxation of $\Theta$,
and integrates as:
$$
G(t,t_w) \equiv \frac{\sigma(t)}{\sigma(t_w)}
\simeq \exp\left[{A\,\left(t_w^\beta-t^\beta\right)}\right]
$$
with
$\beta = 1-\frac{\Delta h'}{\Delta h}$ and
$\quad A=\frac{2\mu\, E_0}{\beta}
\,\left(\frac{E_1}{\Delta h}\right)^{\beta-1}$.
If $\Delta h > \Delta h'$, $\beta>0$ and stress undergoes KWW relaxation. 
The crossover between the short-time exponential relaxation and the long-time
streched exponential is illustrated figure~(\ref{fig:relax1}).
A plateau appears in the crossover region: 
this typical pattern emerges here solely from a change in the dynamical regime of $\Theta$.
Relaxation curves for different ages $t_w$ are shown figure~\ref{fig:relax} (top).
If fitted by $\exp(-t^\beta/\tau(t_w))$, it is easy to check that the apparent 
relaxation timescale $\tau(t_w)$ grows like, $t_w^\alpha$, with $\alpha=1-\beta$; 
this phenomenon, usually refered as sub-aging is here a mere artifact 
of the lin-log representation:
this type of scaling 
-- which are very often used to treat experimental data -- 
may be very misleading and hide, as it does here, the true aging behavior. 
If $\Delta h = \Delta h'$, $\beta=0$, we find $G(t)\simeq 1/t$.
If $\Delta h < \Delta h'$, $\beta<0$ and stress assumes a non-vanishing 
asymptotic value after relaxation, which increases with the age of the sample:
$G(t\to\infty) \simeq \exp\left[{-|A|\,t_w^{-|\beta|}}\right]$.

Another important experimental protocol consist in measuring the 
in-phase and out-of-phase moduli, $G'$ and $G''$, 
when the material is forced  at time $t_w$ by an oscillatory shear, 
$\gamma(t) = \gamma_0\,\sin(\omega\,(t-t_w))$.
In practice, the measurement is performed by integrating
the response a fixed number of periods $n$, for each frequency $\omega$.
Since the material ages, every point on the spectrum depends on 
the whole history of the sample prior to its measurement.
The resulting $G'$ and $G''$ are shown figure~\ref{fig:relax} (bottom). 
They clearly show an $\alpha$ relaxation peak which, 
here is solely the signature of the slow evolution of $\Theta$.
This peak is obtained for large intervals of the parameters,
under the conditions that $\mu$ is somewhat large (which, indeed, 
is reasonable, given the usual values of elastic constants);
it weakens with the increasing age of the material.

To conclude this study, let me consider the regime of stationary deformation 
of the material, at temperature $T$, under a constant shear rate $\dot\gamma$. 
The steady state value of the variable $\Theta$ is the solution of:
$$
\dot\gamma^2 =
E_0\,E_1\,\exp\left[-\frac{\Delta h+\Delta h'}{k\Theta}\right]\,
\left(1-\exp\left(\frac{\delta q}{k\Theta}-\frac{\delta q}{k T}\right)\right)
\quad.
$$
For $\Delta h = \Delta h'$, one finds, 
$\eta\,\dot\gamma=
\sigma(1+\frac{E_0}{E_1}\sigma^2)^{-\frac{\Delta h}{\delta q}}$.
For $\Delta h \ne \Delta h'$, 
and $k\Theta>>\Delta h+\Delta h'$, this solution is dominated by the contribution,
$\Theta=kT/(1+(kT/\delta q)\,\log\left(1-\dot\gamma^2/ (E_1 E_0)\right))$
which, for small $\dot\gamma$, indicates that the system behaves as a newtonian liquid 
with a viscosity, $\eta=\exp\left[\Delta h'/kT\right]/E_0$.
For $kT \ne k\Theta<< \Delta h+\Delta h'$, 
the solution is dominated by its $T\to0$ value, which leads to a power 
relation between stress and strain rate:
$\sigma\sim\dot\gamma^n$, with $n=\frac{\Delta h-\Delta h'}{\Delta h+\Delta h'}$.

This study suggest that, in the low temperature regime, there exists a direct 
relation between the exponent $\beta$ (and $\alpha$) of the KWW relaxation, 
and the power law rheology displayed in steady shear. 
It establishes a relation between two highly dissimilar experimental situations 
-- linear response in the aging regime, and steady shear deformation --
in direct relation to the ratio $\kappa=\Delta h/\Delta h'$ between 
activation barriers associated respectivelly to shear and energy relaxation processes.
$\kappa$ is expected to depend solely on geometrical aspect of the material:
shape of the molecules or polydispersity of colloidal particles.
The existence of such a relation can thus be tested experimentally by 
considering families of closelly related materials in order to allow some 
variation of the exponents.

The equations proposed here are exaggeratelly simple.
Their interest lies in the possibility 
to account for a whole set of properties commonly associated 
with the glass transition, in a over-simplified framework,
in comparison to other studies of rheology.~\cite{sollich97}
Various aspects of these questions certainly deserve further studies. In particular, 
a derivation of the transformation rates would bring out much needed information. 
One may, however, hope that out-of-equilibrium thermodynamics can 
be defined for structural glasses in terms of a few rate equations for 
the appropriate set of thermodynamic quantities.

I thank Jean Carlson and Eric Cl\'ement 
for their support and their interest in my research.
I am especially grateful to Christiane Caroli and Jim Langer 
for their suggestions, and their helpful critiques. 
This work was sponsored by the W. M. Keck Foundation,
and the NSF Grant No. DMR-9813752.


\begin{thebibliography}{20}

\bibitem{falk98}
M.~L. Falk and J.~S. Langer, Phys. Rev. E {\bf 57},  7192  (1998).

\bibitem{falk00}
M.~L. Falk and J.~S. Langer, M.R.S. Bulletin {\bf 25},  40  (2000).

\bibitem{lemaitre02b}
A. Lema\^{\i}tre, Phys. Rev. Lett. {\bf 89},  195503  (2002).

\bibitem{lemaitre02c}
A. Lema\^{\i}tre, A Dynamical Approach to Glassy Materials, cond-mat/0206417,
  2002.

\bibitem{larson99}
R. Larson, {\em The structure and rheology of complex fluids} (Oxford
  University Press, New York, 1999).

\bibitem{turnbull61}
D. Turnbull and M.~H. Cohen, J. Chem. Phys. {\bf 34},  120  (1961).

\bibitem{adam65}
G. Adam and J.~H. Gibbs, J. Chem. Phys. {\bf 43},  139  (1965).

\bibitem{narayanaswamy71}
O.~S. Narayanaswamy, J. Amer. Ceram. Soc. {\bf 54},  491  (1971).

\bibitem{moynihan76}
C.~T. Moynihan {\it et~al.}, Ann. NY Acad. Sci. {\bf 279},  15  (1976).

\bibitem{derec01}
C. Derec, A. Ajdari, and F. Lequeux, Eur. Phys. J. E {\bf 4},  355  (2001).

\bibitem{edwards94}
S. Edwards,  in {\em Granular Matter: An Interdisciplinary Approach}, edited by
  A. Mehta (Springer-Verlag, New-York, 1994), pp.\ 121--140.

\bibitem{struik76}
L.~C.~E. Struik, Ann. NY Acad. Sci. {\bf 279},  78  (1976).

\bibitem{sollich97}
P. Sollich, F. Lequeux, P. H\'ebraud, and M.~E. Cates, Phys. Rev. Lett. {\bf
  78},  2020  (1997).

\bibitem{gallavottibook99}
G. Gallavotti, {\em Statistical Mechanics: a short treatise} (Springer-Verlag,
  Berlin, Heidelberg, 1999).

\end{thebibliography}
\end{document}